\documentclass[showpacs,superscriptaddress,twocolumn,prl]{revtex4} 
\usepackage{amsmath,graphicx}
\usepackage{subfigure}
\usepackage[usenames]{color}
\usepackage[dvipsnames]{xcolor}
\usepackage{ulem}
\usepackage[utf8]{inputenc}
\usepackage[T1]{fontenc}
\usepackage[labelfont=bf,labelsep=space,justification=RaggedRight]{caption}
 
\makeatletter
\renewcommand{\fnum@figure}{Figure \thefigure}
\makeatother

\begin{document} 
 
\title{Evidence for spin-dependent energy transport in a superconductor}
\author{M. Kuzmanović} 
\affiliation{Laboratoire de Physique des Solides (CNRS UMR 8502), Université Paris-Saclay 91405 Orsay, France } 
\author{B. Y. Wu}
\affiliation{Laboratoire de Physique des Solides (CNRS UMR 8502), Université Paris-Saclay 91405 Orsay, France }
\affiliation{Graduate Institute of Applied Physics, National Taiwan University, Taipei 10617, Taiwan } 
\author{M. Weideneder} 
\affiliation{Laboratoire de Physique des Solides (CNRS UMR 8502), Université Paris-Saclay, 91405 Orsay, France } 
\affiliation{Institute for Experimental and Applied Physics, University of Regensburg 93053 Regensburg, Germany } 
\author{C. H. L. Quay}
\affiliation{Laboratoire de Physique des Solides (CNRS UMR 8502), Université Paris-Saclay 91405 Orsay, France } 
\author{M. Aprili} 
\affiliation{Laboratoire de Physique des Solides (CNRS UMR 8502), Université Paris-Saclay 91405 Orsay, France } 
 

\begin{abstract}

In the spin energy excitation mode of normal metals and superconductors, spin up and down electrons (or quasiparticles) carry different heat currents. This mode occurs only when spin up and down energy distribution functions are non-identical, most simply when the two spins have different effective temperatures, and can be excited by spin-polarised current injection into the system. While evidence for spin-dependent heat transport has been observed in a normal metal, these measurements averaged over the distribution function of the electrons. By performing spectroscopy of quasiparticle populations in a mescoscopic superconductor, we reveal distribution functions  which are strongly out-of-equilibrium, i.e. non-Fermi-Dirac. In addition, unlike in normal metals, the spin energy mode in superconductors is associated with a charge imbalance (different numbers of hole- and electron-like quasiparticles) at the superconducting gap edge, in finite Zeeman magnetic fields. Our spectroscopic technique allows us to observe this charge imbalance and thus unambiguously identify the spin energy mode. Our results agree well with theory and contribute to laying the foundation for spin caloritronics with superconductors.

\end{abstract}

\maketitle  

\section{Introduction}

The Seebeck effect, in which a temperature gradient leads to a charge current, was first observed about two centuries ago. Together with its Onsager reciprocal, the Peltier effect, it forms the basis of the field of thermoelectricity or coupled charge and heat transport~\cite{mrs-bulletin}. Coupled charge and spin transport, or spintronics, emerged in the late 1980s~\cite{zutic}. Later, spin caloritronics or coupled heat, charge and spin transport~\cite{boona, bauer} became an experimental reality with the observation of the spin Seebeck effect~\cite{uchida} and spin-dependent Peltier effects~\cite{gravier} in normal metals, and very recently large spin-dependent thermoelectric effects in superconductor-based devices~\cite{machon,ozaeta,kolenda-prl,kolenda-prb}.  

Early work in the field focused on temperature differences between (magnetic) materials associated with spin and/or charge currents. Within a given material, it was pointed out that spin up and down carriers (electrons or quasiparticles) can also have different temperatures~\cite{hatami,heikkila-prb,giazotto,morten,bobkova,heikkila-pss}. When this happens, the spin energy mode of the system is excited and the two spin species carry different heat currents. Evidence for spin-dependent heat transport was recently observed in a normal metal~\cite{dejene} but not in superconductors. Moreover, due to the aggregate nature of the measurements in normal metals (giant magnetoresistance of a spin valve), detailed information on the distribution function could not be obtained. 

Here, we study thin-film superconducting aluminium. As our measurements are spectroscopic, we are able to reveal quasiparticle (QP) populations which cannot be described by effective temperatures (i.e. they are strongly out-of-equilibrium). Instead, they carry an `imprint' of the electron distribution function in the normal metal from which current is injected into the superconductor, to generate QPs. Further, unlike in normal metals, the spin energy mode in superconductors gives rise to a charge imbalance (i.e. different numbers of electron- and hole-like quasiparticles) with a specific energy and magnetic field dependence. Our spectroscopic measurements allow us to observe this imbalance and thus unambiguously identify the spin energy mode. The presence of the spin energy mode in turn necessarily implies that the distribution functions of spin up and down quasiparticles are different.

\section{Spinful Excitation Modes of Out-of-Equilibrium Superconductors}

The ground state of conventional (Bardeen-Cooper-Schrieffer) superconductors is composed of Cooper pairs of electrons in a spin singlet configuration. In equilibrium, this macroscopic quantum state can carry a dissipationless charge current (known as a supercurrent), but not spin or energy currents. In contrast, the single particle excitations, or quasiparticles, are spin-1/2 fermions, which can carry spin, energy and charge currents. The density of states of these QPs ($\rho(E)$) is zero in an energy range $\pm \Delta$ about the Fermi energy ($E_F$), and has coherence peaks just above this gap (Figure 1a).

\begin{figure}
	\label{fig1}
	\includegraphics[width=0.5\textwidth]{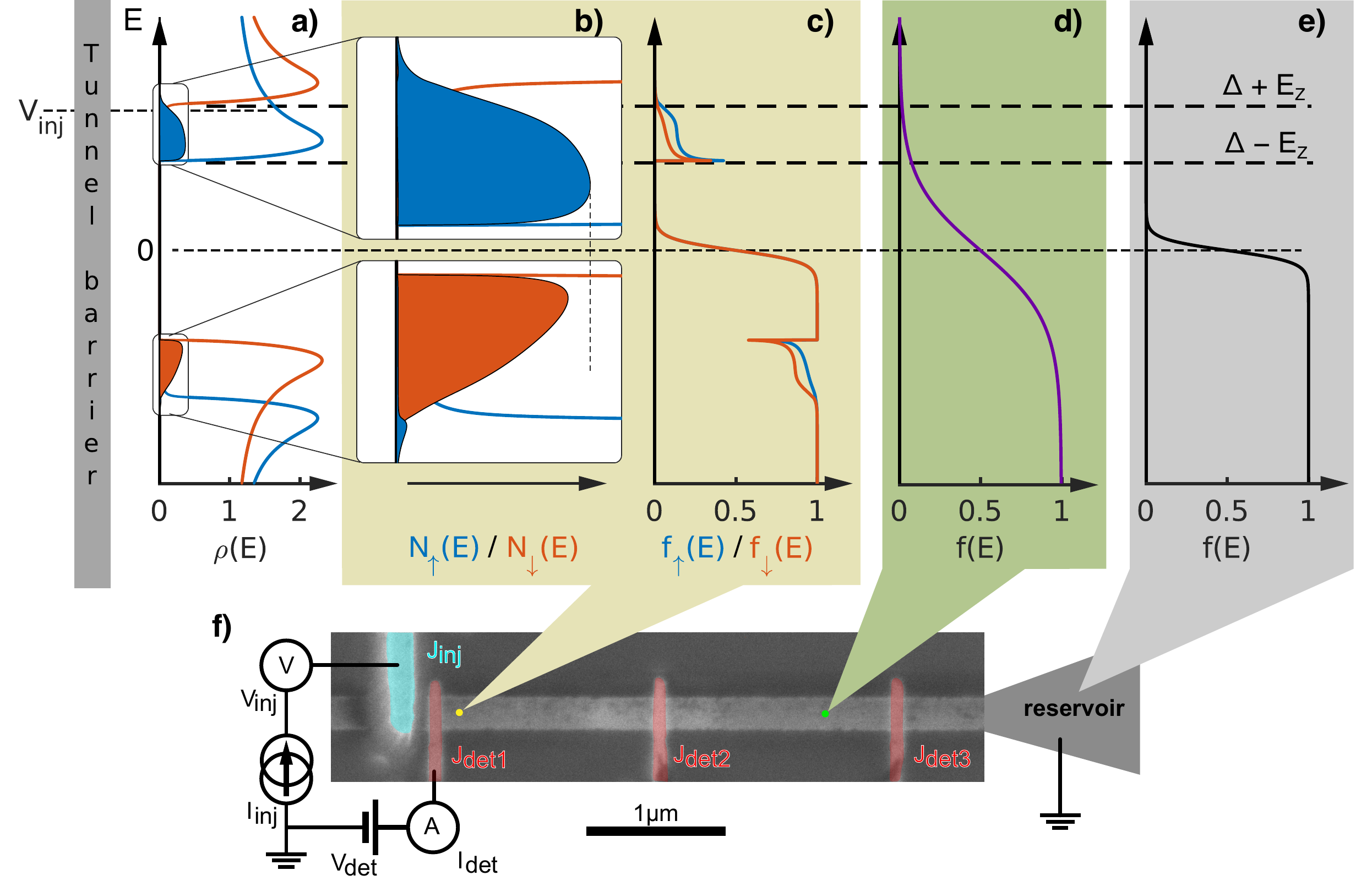}
	\caption{\textbf{$\vert$ Generation and detection of out-of-equilibrium quasiparticles (QP) in a superconductor. 
	a}, Spin up (blue) and down (red) QP density of states (DOS) in the superconductor in an in-plane magnetic field, which induces both a Zeeman splitting and orbital depairing. The blue and red shaded regions are proportional to, respectively, the number of spin up and spin down quasiparticles ($N_\uparrow$ and $N_\downarrow$) near the first detector. This was calculated with the density of states in \textbf{a}, the reservoir distribution function in \textbf{e} and the indicated injection voltage $V_{inj}$. For clarity, the imbalance between the number of electron-like QPs and the number of hole-like QPs (the charge imbalance), has been multiplied five times. This can be seen to occur in a specific energy range. 
	\textbf{b}, Zoom in of \textbf{a}. 
	\textbf{c}, Predicted spin up (blue) and down (red) QP distribution functions at the indicated distance from the injector. The distribution functions show peaks at the superconducting gap edge, as well as a step-like cutoff at $eV_{inj}$.
	\textbf{d}, Farther than an electron-electron interaction length ($\approx$1$\mu$m) from the injector, we expect the quasiparticle distribution function to be spin-independent and close to an effective temperature. The trace shown here is an illustration, not a calculation.
	\textbf{e}, QPs are assumed to be at equilibrium at the reservoir. 
	\textbf{f}, False colour scanning electron micrograph of the device, and a schematic drawing of the spectroscopy measurement setup. The horizontal superconducting wire is 6nm Al. The injector (100nm Cu, cyan) and the detectors (8 nm Al/0.1 nm Pt, red) form tunnel junctions with the wire, with the latter's native oxide as the barrier.}
\end{figure}

Out-of-equilibrium quasiparticle populations in superconductors can be described by the particle energy distribution function $f(E)$. Neglecting the QP spin, $f(E)$ can be decomposed based on symmetry into energy $f_L(E) = f(-E)-f(E)$ and charge $f_T(E) = 1-f(E) - f(-E)$ modes~\cite{schmid,belzig}. The simplest $f(E)$ which excites these modes are, respectively, an effective temperature (different from the lattice temperature) and a charge imbalance. In the presence of a charge imbalance, the number of electron- and hole-like quasiparticles are non-identical, and the quasiparticle chemical potential is different from the Fermi energy.

In the spinful case, this decomposition can be generalised by the addition of spin and spin energy modes, $f_{T3}(E)= f_{T\uparrow}(E)-f_{T\downarrow}(E)$ and $f_{L3}(E)=f_{L\uparrow}(E)-f_{L\downarrow}(E)$~\cite{morten,heikkila-pss}. $f_{L3}$ is most simply excited by a spin-dependent temperature  and $f_{T3}$ by a spin-dependent chemical potential. The spin and spin energy modes only exist if spin up and down QPs have different distribution functions, i.e. if $f_{\uparrow}(E) \neq f_{\downarrow}(E)$. By construction, $f_L$ and $f_{L3}$ are odd in energy, while $f_T$ and $f_{T3}$ are even in energy. In the following, we focus mainly on $f_{L3}$, the spin energy mode.

To generate different spin up and down distribution functions, it is necessary to preferentially excite quasiparticles of one spin species. In thin superconducting films, this can be done by applying an in-plane magnetic field $H$, which lowers (raises) the energy of spin up (down) QPs by the Zeeman energy $E_Z$ and splits the DOS so that only spin up excitations (spin up electron-like and spin down hole-like quasiparticles) are allowed in the energy range $\Delta - E_Z \leq  \vert E \vert \leq \Delta + E_Z$ (Figure 1b). ($E_Z = \mu_{B} H$, with $\mu_B$ the Bohr magneton and $H$ the magnetic field.) Current injection in this energy range thus creates spin-polarised quasiparticles regardless of the magnetic properties of the tunnel barrier or the injector electrode.


For our experiments, we use thin-film superconducting (S) aluminium wires, with a native insulating (I) oxide layer, across which lie normal metal (N) and superconducting (S') electrodes. The former is used as an injector and the latter as detectors (Figure 1f). S is terminated on both sides by reservoirs at a distance of about $5\mathrm{\mu m}$ from the NIS junction. The magnetic field ($H$) is applied in the plane, parallel to S. 

Our basic spectroscopy measurement consists of injecting a constant current $I_{inj}$ at the injector $J_{inj}$, and measuring the current $I_{det}$ and/or the differential conductance $G_{det} = dI_{det}/dV_{det}$ as a function of the applied voltage $(V_{det})$ at one of the detectors ($J_{det1}$, $J_{det2}$ and $J_{det3}$ in Figure 1f). Measurements were performed in a dilution refrigerator with a base temperature of $90\mathrm{mK}$. $J_{det1}$ lies within both a electron-electron interaction length ($\lambda_{e-e} \approx 1 \mu$m~\cite{van-son,santhanam}) and a spin-flip length ($\lambda_{sf} \approx $ 300 nm~\cite{quay-nc}) of the injector.

We model our system using the Usadel-Keldysh equations, which describe out-of-equilibrium diffusive superconductors. (See Supp. Info. for details.) Following Ref.s~\cite{heikkila-pss,bergeret-rmp}, we solve these numerically in one dimension, assuming negligible (inelastic) electron-electron and electron-phonon interactions, and include a Zeeman magnetic field. 
Experimental parameters are used in the model: the normal state diffusion constant $D \approx 10\mathrm{\frac{cm^{2}}{s}}$, $L=10\mathrm{\mu m}$, $R(J_{inj})=13\mathrm{k\Omega}$. The diffusion time from the injector to the reservoirs is $\tau_{diff}=l_{inj-res}^2/D \approx 20\mathrm{ns} $ where $l_{inj-res}$ is the injector-reservoir distance $\approx L/2$. 
As $\tau_{diff}$ is much small than the QP recombination time ($\tau_{rec} \gtrsim 1\mu s$\cite{martinis}), QPs relax and recombine at the reservoirs. At the interface with the injector, the boundary conditions are given by spectral current continuity and the injector distribution function $f_{inj}(E-eV_{inj})$, assumed to be Fermi-Dirac. 


In our numerical results for the closest detector (Figure 1c), we see that the quasiparticle distribution function bears signatures of both the density of states in S (Figure 1b) as well as the distribution function in the injector: It has a peak at $E=\Delta$ and goes sharply to zero at $E = V_{inj}e$. The distribution function is also spin-dependent.

To interpret our experimental results, it is helpful to understand the link between the spin energy mode $f_{L3}$ and charge imbalance by considering the particle number as a function of energy: 
\begin{align}
		\label{qpnumber}
        N(E) & = N_\uparrow(E) + N_\downarrow(E) = f_\uparrow(E)\rho_\uparrow(E)+f_\downarrow(E)\rho_\downarrow(E)\\
            & = \rho_+(E)[1-f_L(E)-f_T(E)]-\rho_-[f_{T3}(E)+f_{L3}(E)]
\end{align}
\noindent Here $\rho_\uparrow(E)$ and $\rho_\downarrow(E)$ are the DOS of spin up and spin down QPs respectively, $\rho_+(E) := \tfrac{1}{2}[\rho_\uparrow(E)$ + $\rho_\downarrow(E)]=\rho(E)$ and $\rho_-(E) \equiv \tfrac{1}{2}[\rho_\uparrow(E)$ - $\rho_\downarrow(E)]$.

Here we notice that the term $\rho_-(E)f_{L3}(E)$ is even in energy, which means that the spin energy mode $f_{L3}$ \textit{adds} particles at both positive and negative energies, and raises the overall quasiparticle chemical potential, thus creating a charge imbalance. (Figure 1b) In addition, the multiplication by $\rho_-(E)$ means that $f_{L3}$ add particles in the energy range $\Delta - E_Z \leq \vert E \vert \leq \Delta + E_Z$, regardless of the injection voltage or other experimental parameters. (Figure 1b) $f_{T}$ also creates a charge imbalance, which however appears at low magnetic fields and high energies. Our spectroscopic technique allows us distinguish between $f_{L3}$ and $f_{T}$, based on their different energy dependences. We refer the reader to Ref.~\cite{heikkila-pss} and the Supp. Info. for further theoretical details.

\section{Spectroscopic Spin-sensitive Quasiparticle Detection}

We first characterise both injector and detector junctions, and explain our spectroscopy technique. Figure 2a shows the differential conductance of the injector $G_{inj} = dI_{inj}/dV_{inj}$ as a function of the applied voltage $(V_{inj})$ at different $H$. At the temperatures of our experiment, $G_{inj}$ is almost exactly proportional to the density of states in S~\cite{tinkham}. We can see that $H$ induces Zeeman splitting of the QP DOS. $H$ also couples to the orbital degree of freedom, inducing screening supercurrents and hence a rounding of the QP coherence peak due to orbital depairing~\cite{tinkham, fulde}. The depairing parameter, found by fitting the data with an Abrikosov-Gor'kov depairing (see Supp. Info.), is $\alpha = R_{ORB} H^2$, with $R_{ORB} \approx 6.5\mathrm{\frac{\mu eV}{T^2}}$, and the critical field $H_c \approx 2.7\mathrm{T}$. In the results shown here, the Zeeman energy is always greater than the depairing parameter. (See Supp. Info. for details.) 

\begin{figure}
	\label{fig2}
	\includegraphics[width=0.35\textwidth]{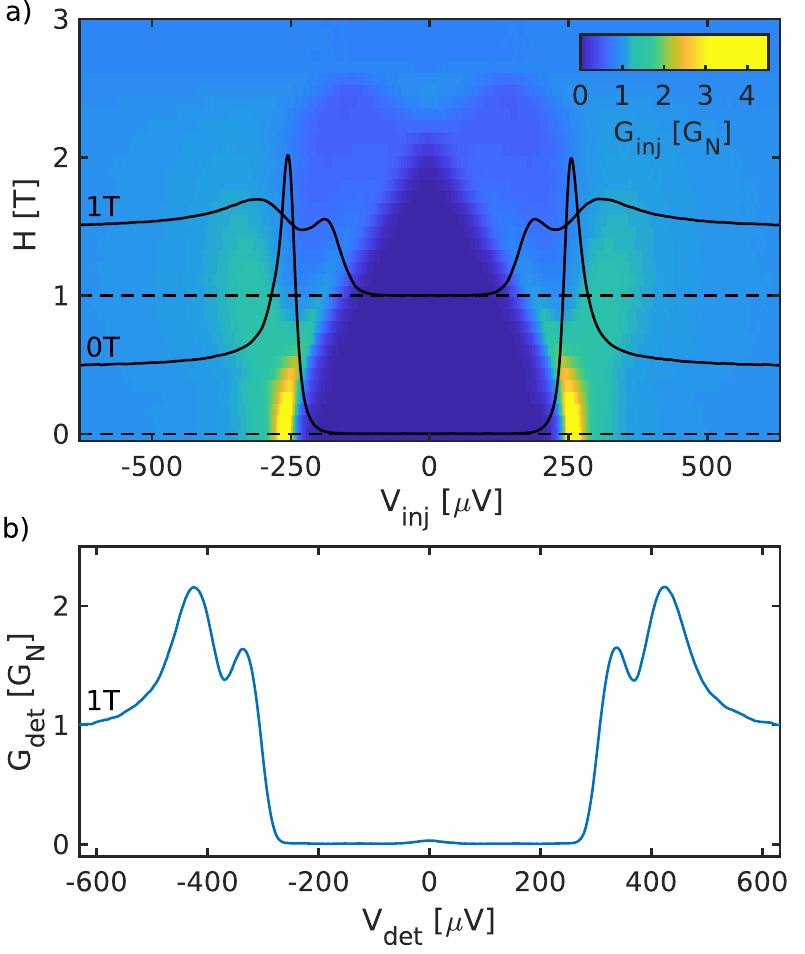}
	\caption{\textbf{$\vert$ Characterisation of injector and detector junctions. a}, Differential conductance of the injector junction $G_{inj}$ as a function of injector voltage $V_{inj}$ and magnetic field $H$, and slices at $H=0$T and $H=1$T (black traces). \textbf{b}, Differential conductance of the detector junction $G_{det}$ as a function of the detector voltage $V_{det}$ at $H=1\mathrm{T}$ without any injection current. We see the Zeeman splitting of the quasiparticle density of states in the superconducting wire as the detector is not Zeeman-split.}
\end{figure}

If the detector temperature is much smaller than the superconducting energy gap in S' ($k_B T_{det} \ll \Delta_{det}$, with $k_B$ Boltzmann's constant), the differential conductance of SIS' junctions as a function of the applied voltage in the subgap region $V < (\Delta + \Delta_{det})/e $ is given by 

\begin{equation} 
\label{GV_SIS}
G_{det}(V_{det})= \frac{1}{e R_{\mathrm{N}}}\int \rho(E)f(E) \frac{\partial \rho_{det}(E+eV_{det})}{\partial V_{det}}dE 
\end{equation}

\noindent where $\rho_{det}(E)$ the density of states in S', $e$ the electron charge and $R_N$ the normal state resistance of the detector junction.

Most of the integral comes from the coherence peak in $\rho_{det}$ at $E=\Delta_{det}$. This peak picks out the number of quasiparticles in S ($\rho(E)f(E)$), shifted by $\Delta_{det}$. In other words, $G_{det}(V_{det}-\Delta_{det}/e)$ gives the number of QPs at energy $E = eV_{det}$, while $I_{det}(V_{det}-\Delta_{det}/e)$ gives the total number of QPs for $E \leq eV_{det}$. Our measurements thus give us spectroscopic information on the QPs. (See Supp. Info. for details.) 

At finite magnetic fields, these spectroscopic measurements become spin-sensitive if Zeeman spin-splitting occurs in S but not in S'; the unsplit coherence peak in S' separately probes the number of QPs in S at the two gap edges for spins up and down, respectively at $V_{det}^{\uparrow (\downarrow)} = |\Delta \pm \mu_B H - \Delta_{det}|/e$. 

We suppress the spin-splitting in S' through the strong spin-orbit coupling of sprinkled Pt, which acts as a spin-mixer. (See Methods, Supp. Info. and as Ref.s~\cite{tedrow,meservey,bruno,fulde}) Figure 2b shows $G_{det}(V_{det})$ at different $H$ and $I_{inj}=0$. At $H=1\mathrm{T}$, we see two peaks, as expected for a non-spin-split detector. (Were there a Zeeman splitting in S' equal to that in S, the situation would be equivalent to two SIS junctions in parallel, one for each spin, and there would be a single peak in $G_{det}(V_{det})$ instead of two~\footnote{The asymmetrical signal in Figure 5 would remain in the data, but we would be unable to differentiate the contribution from the two spins and clearly identify $f_{L3}$.}.) We note also that the detector current is typically $0.1 - 1\mathrm{nA} \ll I_ {inj} \sim10-100\mathrm{nA}$ throughout the subgap region: the detector is close to equilibrium

\section{Non-Fermi-Dirac Quasiparticle Energy Distributions}

Measurements at zero magnetic field already reveal non-Fermi Dirac distributions. Figure 3a shows the current-voltage characteristics of the closest detector junction at two injection currents: 0nA (black trace) and 120nA (red trace). We focus on the low-voltage range before the abrupt rise of $I_{det}$ at $V_{det}=(\Delta+\Delta_{det})/e$, where the opposite-energy coherence peaks of S and S' align. We see that the red trace is higher than the black. This indicates the presence of additional QPs created by injection~\footnote{Such measurements of `excess QPs' have been made in extended junctions, but because of the spatial averating, the spectroscopic information was lost.}.

\begin{figure}
	\label{fig3}
	\includegraphics[width=0.5\textwidth]{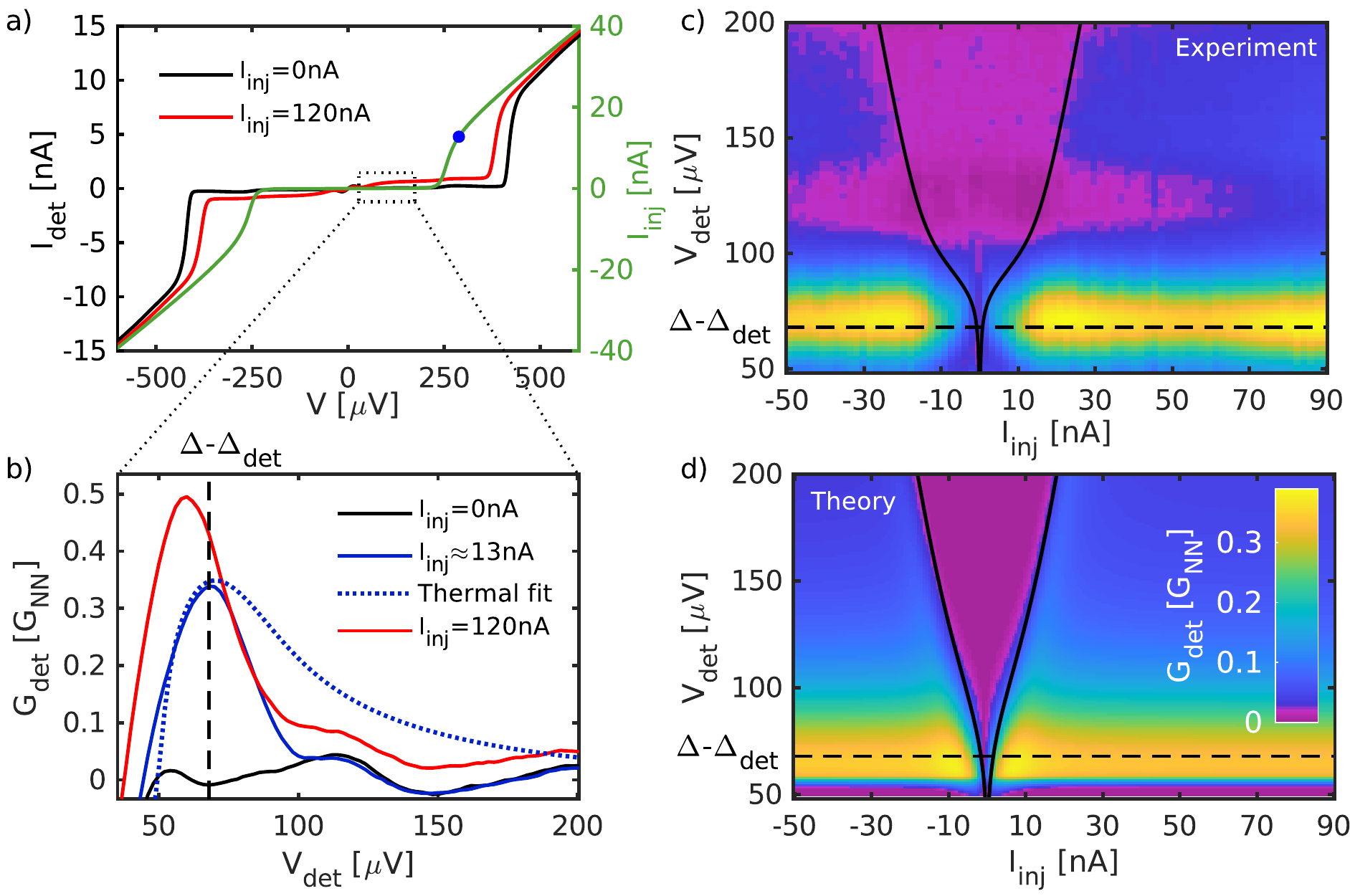}
	\caption{\textbf{$\vert$ Non-Fermi-Dirac quasiparticle distribution. 
	a}, Current $I_{det}$ as a function of voltage $V_{det}$ across the SIS' detector junction $J_{det1}$ for injection currents $I_{inj}=0\mathrm{nA}$ (black) and $I_{inj}=120\mathrm{nA}$ (red). On the right vertical scale, $I_{inj}$ as a function of voltage $V_{inj}$ across the NIS injector junction $J_{inj}$ (green). $H=0$ throughout this figure.
	\textbf{b}, Differential conductance $G_{det}$ as a function of $V_{det}$ across $J_{det1}$ for $I_{inj}=0\mathrm{nA}$ (black), $I_{inj}\approx13\mathrm{nA}$ (blue, blue dot in \textbf{a}), and $I_{inj}=120\mathrm{nA}$ (red). The vertical dashed line indicates $eV_{det}=\Delta - \Delta_{det}$; $G_{det}$ at this voltage is proportional to the number of quasiparticles in the superconducting wire at $E=\Delta$. An attempted fit with an effective temperature $T^*\approx1.1\mathrm{K}$ in S reproduces the peak at $I_{inj}=13\mathrm{nA}$, but grossly overestimates the QP population at higher energies (dashed blue line). In this fint, we use the experimentally determined values $\Delta=245\mathrm{\mu eV}$ and $\Delta_{det}=180\mathrm{\mu eV}$, $T_{det}=90\mathrm{mK}$ and a phenomenological depairing $\alpha \approx 1\% \Delta$.
	\textbf{c}, $G_{det}$ at $J_{det1}$ as a function of $V_{det}$ and $I_{inj}$ with the slice at $I_{inj}=0$ subtracted from all data. The black lines show the measurement of $\pm I_{inj}(V_{inj})$ from \textbf{a} shifted downwards by $\Delta_{det}/e$. The black lines fall at the location of a step-like feature in the colour map, as expected: as shown in Figure 1b, QPs in S are created up $E\approx eV_{inj} + k_B T$, leading to a step-like cutoff in the distribution function. The dashed line again indicates $eV_{det}=\Delta - \Delta_{det}$, where the QP density is maximal due to the coherence peak in the DOS of S. 
	\textbf{d}, Theoretical prediction of \textbf{c}, with the $\Delta$, $\Delta_{det}$ and $\alpha$ as in \textbf{b}}
\end{figure}

This creation of quasiparticles by current injection can also be seen in the differential conductance measurement, $G_{det}(V_{det})$ at three values of $I_{inj}$: 0nA, $\approx$ 13nA and 120nA (Figure 3b). Here, we see more clearly that most of the quasiparticles are at the gap edge ($eV_{det}=\Delta$). If we try to fit the trace at $I_{inj}\approx 13$nA with a thermal QP distribution, it is clear that this grossly over-estimates the number of QPs at high energies (Figure 3b, dotted line). The quasiparticles do not thermalise. 

Instead, as shown in our calculations (Figure 1) and discussed earlier, the quasiparticle states in S are filled up to $V_{inj}$: the electron distribution function in N is `imprinted' onto the quasiparticles in S. This can be seen by overlaying the $I_{inj}(V_{inj})$ measurement in Figure 3a, shifted by $\Delta_{det}/e$, onto a plot of $G_{det}$ as a function of $(V_{det})$ and $I_{inj}$ (Figure 3c). Note that, at each current, the injector voltage falls exactly at the location of a step in $G_{det}$ (seen here as a change in colour). The accumulation of quasiparticles at the gap edge in S can also be seen on this colour scale as a yellow horizontal feature.

Our calculations reproduce both the step-like feature corresponding to $I_{inj}(V_{inj}+\Delta_{det}/e)$, as well as the horizontal feature (Figure 3d). Thus, at a distance of about $300\mathrm{nm} \ll \lambda_{e-e}$ from the injector (i.e. at $J_{det1}$) and in the energy range of interest for the detection of the $f_{L3}$ mode, the quasiparticles have not yet thermalised, and it is reasonable to neglect electron-electron interactions.

\section{Spin Energy Mode}

At finite magnetic fields, current injection at low energies becomes spin-polarised: we expect different distribution functions for spin up and down quasiparticles, and in particular to excite the spin energy mode. We show in Figure 4a calculations of $G_{det}$ as a function of $V_{det}$ (in the sub-gap region) and of $I_{inj}$, at $1\mathrm{T}$ where the density of states in S is well spin-split (Figure 2a). Following features from low to high energies, we expect peaks in $G_{det}(V_{det})$ at $eV_{det}=(\pm|\Delta-\Delta_{det}-E_Z|)$ which we shall call $P_2$ and $P_3$, corresponding to the coherence peaks of spin up excitations (spin up electron-like or spin down hole-like quasiparticles). Peaks at $V_{det}=\pm|\Delta-\Delta_{D}+E_Z|$ ($P_1$ and $P_4$), corresponding to the coherence peaks of spin down excitations, appear when $I_{inj}$ is increased and spin down QPs are also injected.

\begin{figure}
	\label{fig4}
	\includegraphics[width=0.5\textwidth]{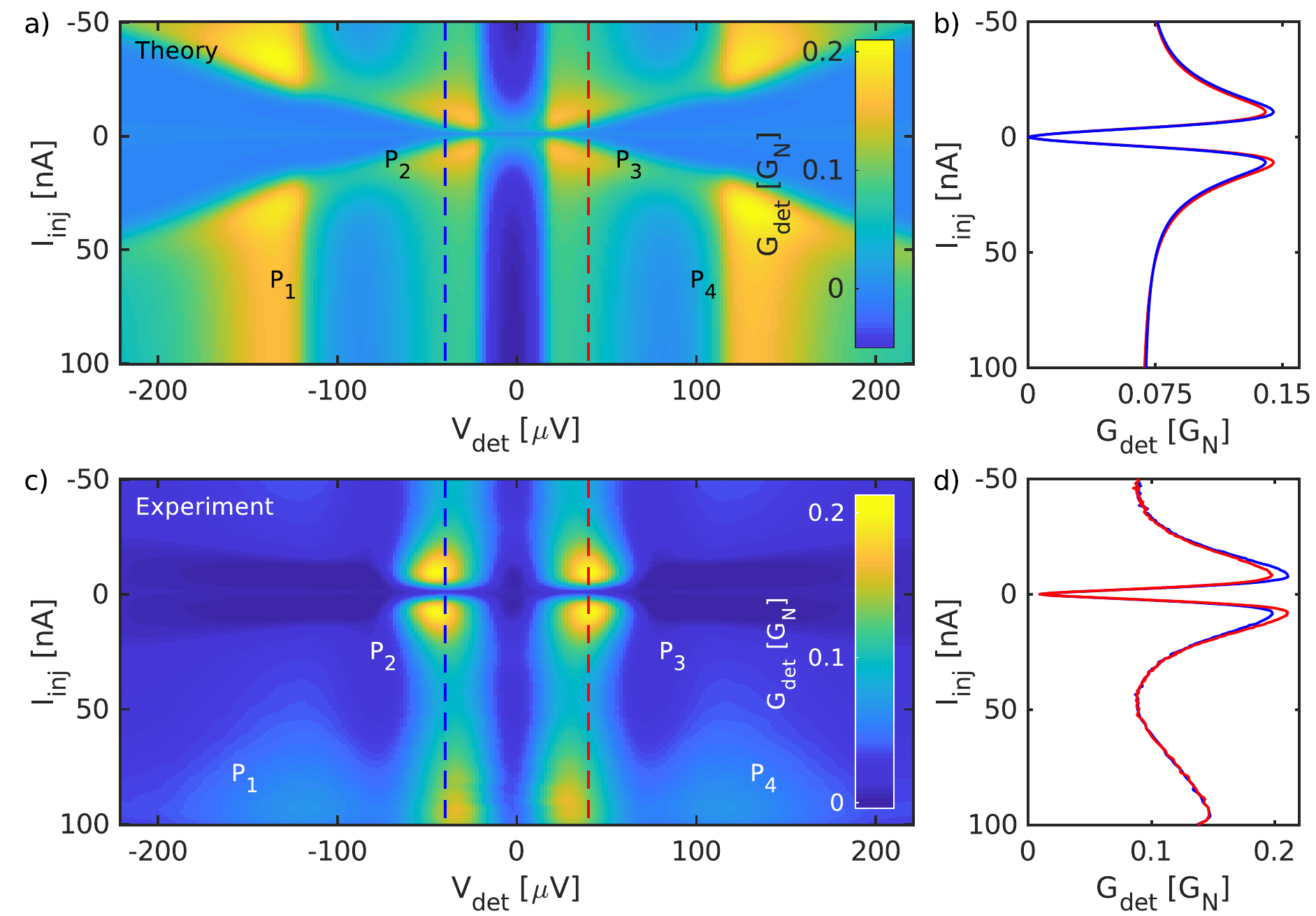}
	\caption{\textbf{$\vert$ Spin energy mode. a,c}, Theoretical calculations for and measurements of the differential conductance as a function of voltage and injection current at $J_{det1}$ for $H=1$T. The peaks P$_1$ -- P$_4$ observed experimentally and reproduced in our calculations are due to spin up (P$_2$, P$_3$) and spin down (P$_1$, P$_4$) excitations. \textbf{b,d}, Vertical slices of \textbf{a} and \textbf{b} at $eV_{det}=\pm|\Delta-\Delta_{det}-\mu_{B} H|$ (red for + and blue for -), indicated by the dashed blue and red lines. A charge imbalance can be seen, i.e. the red and blue traces are not identical.}
\end{figure}

Comparing this to the data (Figure 4c), we see $P_2$ and $P_3$ clearly, but $P_1$ and $P_4$ are less prominent. This is due to the increased electron-electron interaction at high energies and QP number. (For clarity, the Josephson (i.e. supercurrent) contribution has been subtracted from $G_{det}$. See Supp. Info. for details.)

Next, we compare the number of electron- and hole-like quasiparticles by taking two slices of Figure 4c at $eV_{det}=+|\Delta-\Delta_{D}-\mu_{B} H|$ (Figure 4d). The traces are not identical. The difference between them, which is the charge imbalance, is maximal at $I_{inj} \approx 8$nA, corresponding to maximal spin polarisation of the injection current, i.e. when the injection voltage is just below the coherence peak of the second spin species. This charge imbalance is also reproduced in the calculation (Figure 4b). 

The charge imbalance associated with $f_{L3}$ has particular energy and magnetic signatures: it is expected to appear in the energy range $\Delta - E_Z \leq  \vert E \vert \leq \Delta + E_Z$. In Figure 5a, we plot the component of the data in Figure 4a which is odd in $V_{det}$, which gives the charge imbalance. The odd component is indeed largest in the expected energy range. As the magnetic field is decreased, the charge imbalance is reduced, also as expected for the spin energy mode (Figure 5b): it is zero at zero magnetic field, and becomes visible when $E_Z > 3.5k_B T$. At $H = 1 \mathrm{T}$. Our calculations reproduce the data well (Figure 5b, dash-dotted line). 

\begin{figure}
	\label{fig5}
	\includegraphics[width=0.5\textwidth]{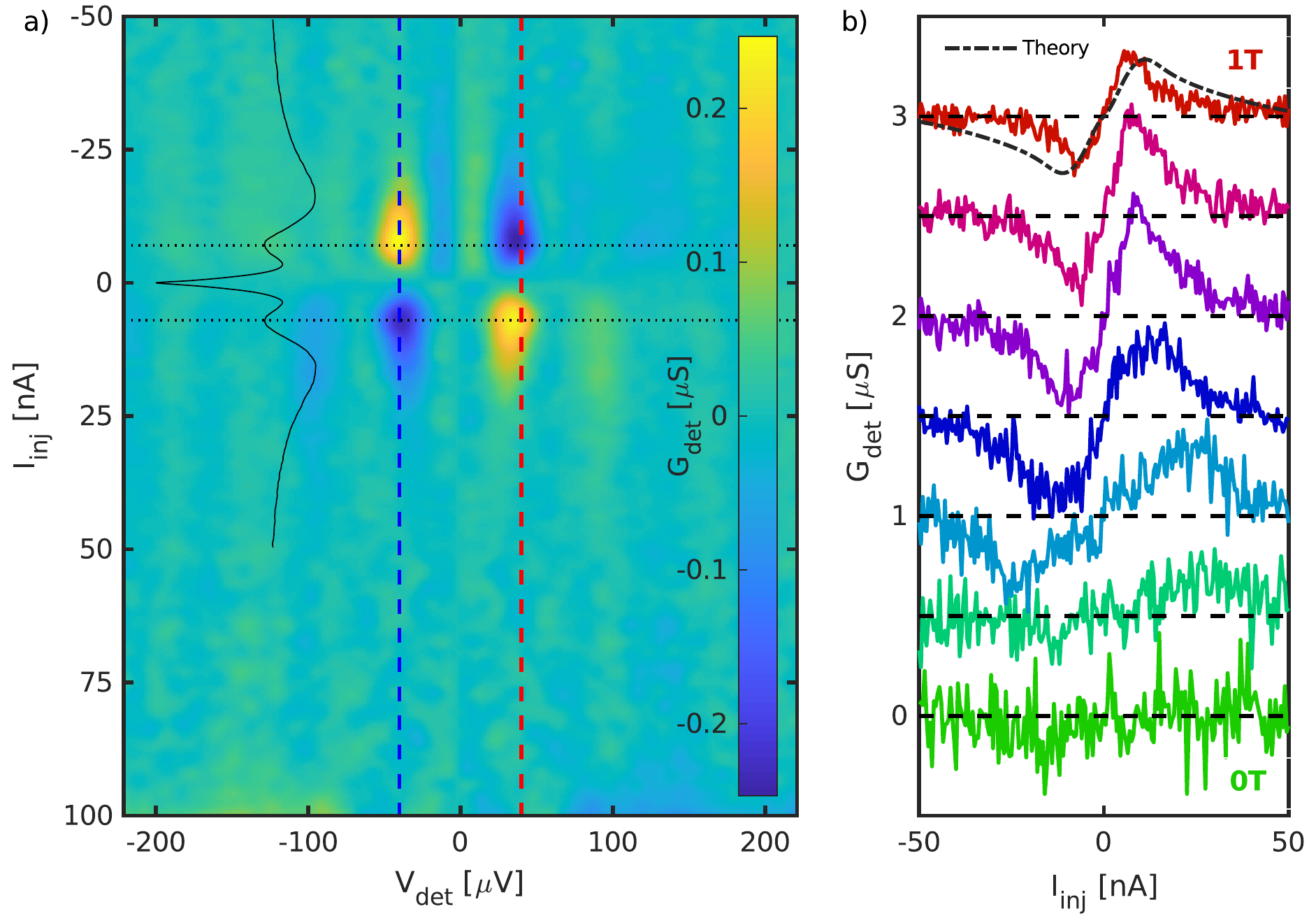}
	\caption{\textbf{$\vert$ Close-up of the spin energy mode. a}, The odd-in-energy component of Figure 4c, corresponding to a charge imbalance. This only appears at the gap edge: the vertical dashed lines indicate $V_{det}=\pm|\Delta-\Delta_{det}-E_Z|/e$. The signal is maximal (horizontal dotted lines) when only spins of one species are injected: $G_{inj}(V_{inj})$ is shown (thin black line) on the left and top axes. \textbf{b}, A vertical slice of \textbf{a} at $V_{det}=(\Delta-\Delta_{det}-E_Z)/e$ and the same measurement at different magnetic fields linearly spaced down to $H=0\mathrm{T}$. The theoretical prediction for $H=1\mathrm{T}$ is shown in black. The charge imbalance increases slightly then decreases as the magnetic field is lowered. At $H=0$ it is undetectable.}
\end{figure}

The odd component of the data in Figures 4b and 4d, which comes from $f_{L3}$, is small compared to the even component, which comes from either $f_{L}$ or $f_{T3}$. The quasiparticles from $f_{L}$ or $f_{T3}$ contribute to a finite magnetisation in the superconductor, previously detected by other methods~\cite{hubler,silaev,bobkova-long-range,bobkova-prb}. At $H=0$, we recover the previously observed charge imbalance signal~\cite{lemberger,hubler-charge,kleine,cadden-zimansky,takane}, associated with the $f_T$ mode, which occurs at high energies and low magnetic fields. (See Supp. Info.)


As expected, we do not observe $f_{L3}$ at $J_{det2}$ or $J_{det3}$, where the spin up and down QP distribution functions have become identical. (See Supp. Info.)

Compared to normal metals and semiconductors, the spin energy mode in superconductors has the advantage of being excitable by using the spin-split DOS. Its association with an energy-localised charge imbalance make it easy to distinguish from other modes. Using superconductors as detectors allowed us to have spectroscopic information on the quasiparticles, by using the coherence peak in the detector density of states. This work paves the way for new spin-dependent heat transport experiments, as well as the generation of spin supercurrents by out-of-equilibrium distribution functions in conventional superconductors~\cite{heikkila-pss,aikebaier}.


\bibliographystyle{naturemag}
\bibliography{references}

\section{Acknowledgements}

\noindent We acknowledge valuable discussions with Tero Heikkilä, Mikhail Silaev and Wolfgang Belzig; and an ANR JCJC grant (SPINOES) from the French Agence Nationale de Recherche. BYW is grateful for a  College of Science (CoS) Travel Grant and Scholarship from the National Taiwan University. We also thank Freek Massee, Hadar Steinberg and Suchitra Sebastian for helpful comments on the manuscript.

\section{Author Contributions} 

\noindent MK fabricated the devices and performed the numerical calculations. MK and MA made the measurements. MK, MA and CQHL analysed the data and wrote the manuscript. BYW and MW were involved in earlier stages of the work.

\section{Competing Financial Interests}

\noindent The authors declare no competing financial interests.

\section{Methods}

\noindent The superconducting wire is 6nm Al, while the injector is 100nm Cu and the detectors 8nm Al/0.1nm Pt. The devices were fabricated with standard electron-beam lithography and evaporation techniques. The NIS and SIS' junctions have conductances per unit area $\approx 1.9 \mathrm{\frac{mS}{\mu m^2}}$ and $\approx 3.3 \mathrm{\frac{mS}{\mu m^2}}$ respectively (corresponding to barrier transparencies of $\approx 2\times 10^{-5}$). All measurements were performed using standard lock-in techniques in a dilution refrigerator with a base temperature of 90mK. The lock-in frequency is typically $17-37\mathrm{Hz}$ and the excitation voltage  $5\mathrm{\mu V}$. The out-of-plane component of $H$ was compensated to be $\leq 1\%$ of the total field.

\end{document}